\documentclass{Interspeech}

\interspeechcameraready

\usepackage{amsmath,amsfonts}
\usepackage{algorithmic}
\usepackage{array}
\usepackage{textcomp}
\usepackage{stfloats}
\usepackage{url}
\usepackage{verbatim}
\usepackage{graphicx}
\usepackage{multirow}
\usepackage{booktabs}
\usepackage{makecell}
\usepackage{xcolor}
\usepackage{hyperref}
\usepackage{subcaption}

\usepackage{lipsum}

\makeatletter
\newcommand\footnoteref[1]{\protected@xdef\@thefnmark{\ref{#1}}\@footnotemark}
\makeatother

\title{Neurodyne: Neural Pitch Manipulation with Representation Learning \\ and Cycle-Consistency GAN}

\author[affiliation={1,2}]{Yicheng}{Gu}
\author[affiliation={2}]{Chaoren}{Wang}
\author[affiliation={2}]{Zhizheng}{Wu}
\author[affiliation={1}]{Lauri}{Juvela}

\affiliation{Acoustic Lab}{Aalto University}{Espoo, Finland}
\affiliation{School of Data Science}{The Chinese University of Hong Kong, Shenzhen}{China}
\email{yicheng.gu@aalto.fi}
\keywords{speech recognition, human-computer interaction, computational paralinguistics}

\keywords{Pitch manipulation, singing voice synthesis, generative adversarial networks (GAN), representation learning}

\usepackage{comment}

\begin{document}

\maketitle

\begin{abstract}

Pitch manipulation is the process of producers adjusting the pitch of an audio segment to a specific key and intonation, which is essential in music production. Neural-network-based pitch-manipulation systems have been popular in recent years due to their superior synthesis quality compared to classical DSP methods. However, their performance is still limited due to their inaccurate feature disentanglement using source-filter models and the lack of paired in- and out-of-tune training data. This work proposes Neurodyne to address these issues. Specifically, Neurodyne uses adversarial representation learning to learn a pitch-independent latent representation to avoid inaccurate disentanglement and cycle-consistency training to create paired training data implicitly. Experimental results on global-key and template-based pitch manipulation demonstrate the effectiveness of the proposed system, marking improved synthesis quality while maintaining the original singer identity.

\end{abstract}

\section{Introduction}

Pitch manipulation 
is an essential process in music production where the producer can adjust the pitch of an audio segment to correct out-of-tune notes and improve the intonation. However, the performance of existing pitch-manipulation systems is still limited, which will generate unnatural modified audio with audible artifacts.
Thus, building a system that can generate high-fidelity audio given the modified pitch contour is needed.

Classical DSP-based methods can be classified into signal and parametric modification systems. 
Signal modification systems, such as TD-PSOLA~\cite{psola}, relocate pitch periods in the time domain, whereas 
parametric modification systems, such as WORLD~\cite{world}, decompose the signal into acoustic parameters and resynthesize the signal from the modified ones.
%
However, due to inaccuracies in DSP modeling, such systems often have metallic artifacts and timbre inconsistency in the modified audio. 
Recent advances in deep learning (DL)-based systems offer improved audio quality.
Specifically,~\cite{popbutfy} adapts a two-stage generation scheme conditioned on modified pitch contours, followed by DiffPitcher~\cite{diffpitcher} using a diffusion-based architecture.
These systems operate solely on acoustic features and lack audio-level optimization, which limits their synthesis quality. In contrast,
DeepAutotuner~\cite{deepautotuner} and KaraTuner~\cite{karatuner} generate audio in an end-to-end way, followed by SiFiGAN~\cite{sifigan} and HarmonicNet~\cite{pcnsf} adapt the neural vocoders~\cite{HiFiGAN, bigvgan} for decoding, obtaining state-of-the-art (SOTA) performance. To get a faster inference speed, PeriodGrad~\cite{periodgrad} applies a diffusion model on the time-domain signal, and FIRNet~\cite{firnet} utilizes a harmonic-plus-noise synthesizer to reduce the model size.

Two central problems remain with the existing systems.
First, current models \cite{diffpitcher, sifigan, pcnsf} rely on a source-filter decomposition \cite{world} for disentanglement, and assume independence between pitch and spectral envelope. However, the independence is not complete, which can lead to timbre inconsistency in modified audio.
Second, ground truth paired data with in- and out-of-tune examples are unavailable, significantly limiting the system's generalization ability, especially in extreme modifications.

To address these issues, we propose \textit{Neurodyne}. Instead of using inaccurately disentangled features, our system employs adversarial representation learning to learn a pitch-independent latent representation, following~\cite{hfc} and~\cite{ns3}. 
Moreover, we apply two kinds of cycle-consistency training schemes, including the regular inversion cycle-consistency~\cite{cyclegan}, and a novel composition cycle-consistency. 
This allows us to create paired data implicitly based on three widely used pitch-manipulation operations in real-world music production (key, variance, and transient manipulation), making our model more robust in different scenarios. 
Experiment results in both global-key~\cite{sifigan} and template-based~\cite{diffpitcher} pitch-manipulation confirm the effectiveness of our system, which outperforms the baselines in audio quality, singer similarity, and pitch accuracy.


\section{Methodology}

\begin{figure*}[t]
    \centering
    \begin{subfigure}[b]{0.66\textwidth}
    \centering
    \includegraphics[width=\textwidth]{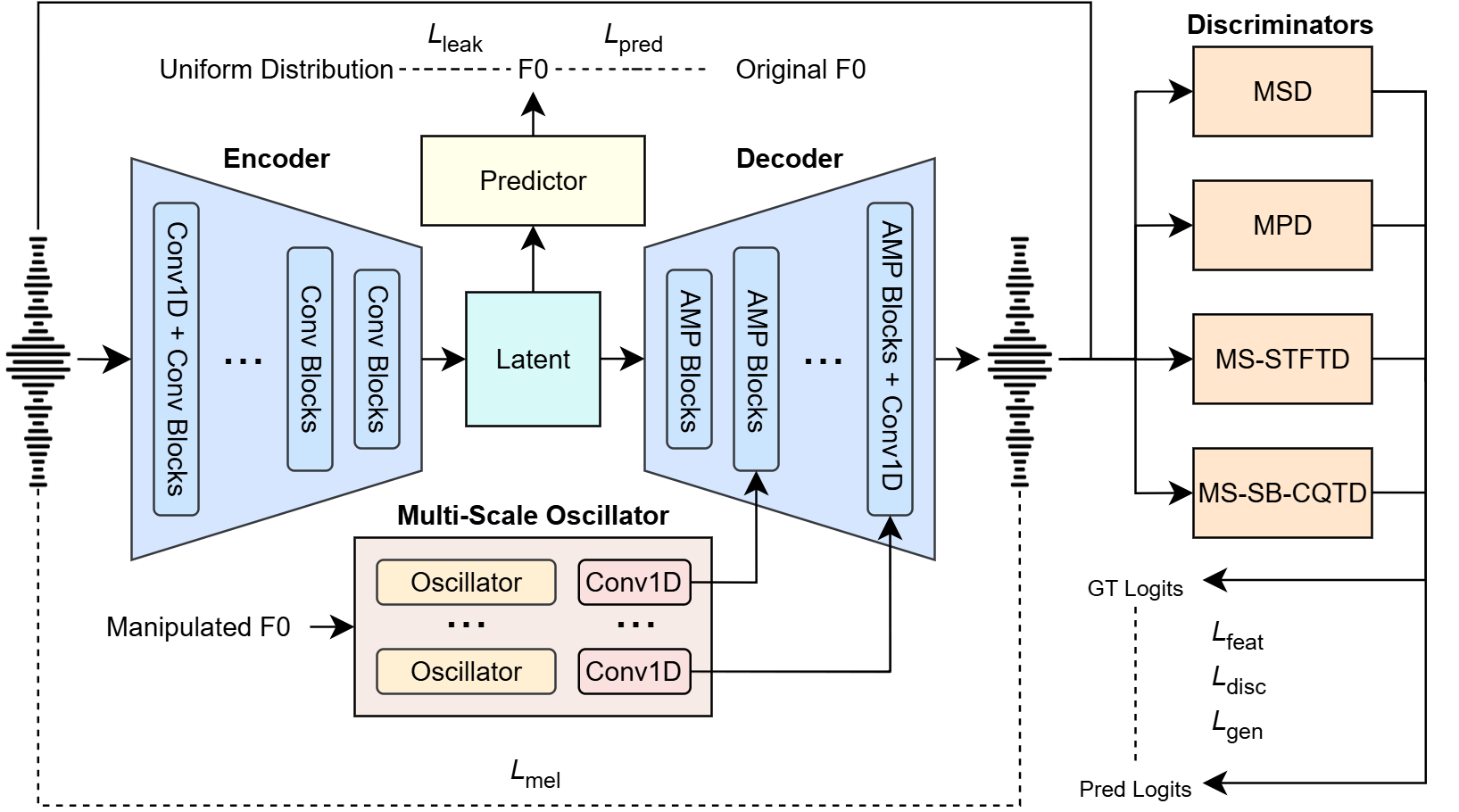}
     \caption{Neurodyne Architecture
     }
     \label{fig:model}
    \end{subfigure} 
    \hfill
    \begin{subfigure}[b]{0.33\textwidth}
    \centering
    \includegraphics[width=\textwidth]{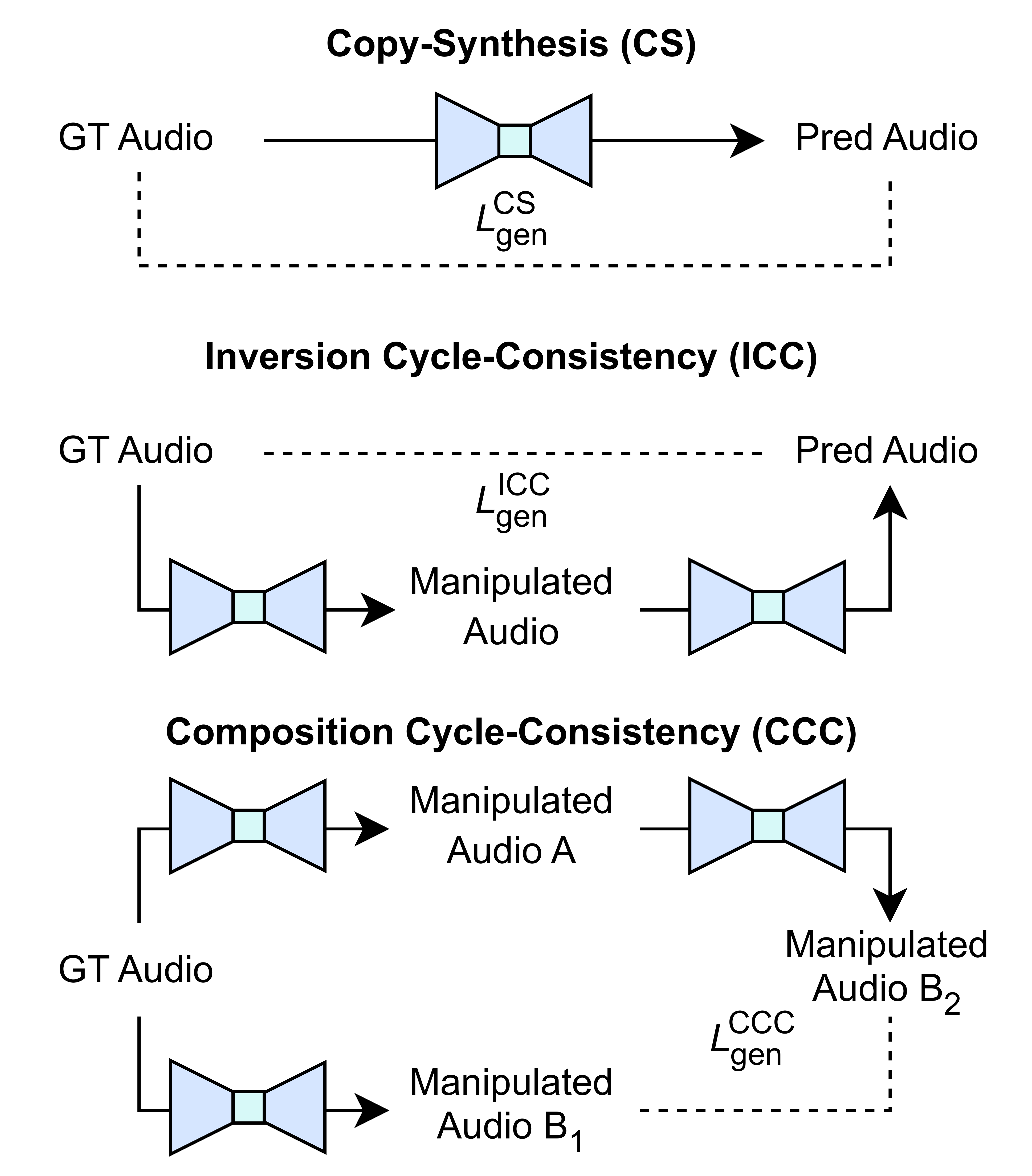}
     \caption{Training Scheme
     }
     \label{fig:train}
    \end{subfigure} 
    \caption{Architecture and training schemes for Neurodyne. The model consists of an encoder, a pitch predictor, a decoder, a multi-scale oscillator, and four different discriminators. The training scheme adopts inversion and composition cycle-consistency training to construct paired in- and out-of-tune data implicitly.}
    \label{fig:overall}
\end{figure*}

This section details the model architecture, training scheme, and pitch-manipulation strategies of \textit{Neurodyne}.

\subsection{Model Architecture}

As shown in Fig.~\ref{fig:model}, Neurodyne consists of an encoder, a pitch predictor, a decoder, a multi-scale oscillator, and four different discriminators. The encoder includes an initial Conv1D layer and five CNN-based Resblocks following DAC~\cite{dac}. The pitch predictor consists of a CNN block with a Conformer~\cite{conformer} encoder, following TorchFCPE\footnoteref{footnote:fcpe}. The decoder includes five CNN-based Resblocks with the oversampling technique applied on the activation function to alleviate aliasing effects following BigVGAN~\cite{bigvgan} and a final Conv1D layer. The multi-scale oscillator consists of four NSF~\cite{nsf}-based excitation oscillators in different sampling rates followed by four Conv1D layers, where the pitch information will first be converted to excitations in different resolutions and added to the intermediate features after the upsampling layer via the Conv1D layer to avoid aliasing effects in the conditional embedding.
We mix time-domain and time-frequency-representation-based discriminators following~\cite{cqtjournal} to obtain a better synthesis quality without sacrificing the inference speed, which are: Multi-Period Discriminator (MPD)~\cite{HiFiGAN}, Multi-Scale Discriminator (MSD)~\cite{HiFiGAN}, Multi-Scale STFT Discriminator (MS-STFTD)~\cite{encodec}, and Multi-Scale Sub-Band CQT Discriminator (MS-SB-CQTD)~\cite{cqt}.  

\subsection{Adversarial Representation Learning}

We apply adversarial representation learning to obtain a pitch-independent latent representation following the Hider-Finder-Combiner (HFC)~\cite{hfc} approach. Specifically, the pitch predictor aims to extract the pitch information from the latent representation. In contrast, the encoder aims to avoid pitch information leakage to the latent representation, giving:
\begin{equation}
\begin{split}
    &L_\text{pitch}(y, \hat{y}) = L_\text{BCE}(y, \hat{y}), \\
    &L_\text{leak}(\hat{y}) = n_\text{bins} ^ {2 / (n_\text{bins} - 1)}\text{Var}(\hat{y}), \\
\end{split}
\end{equation}
where $n_\text{bins}$ is the number of pitch bins, $y$ is the ground truth and $\hat{y}$ is the predicted pitch vector following CREPE~\cite{crepe}.



\subsection{Cycle-Consistency Training}

The proposed cycle-consistency training scheme consists of three complementary training goals, as described below. 

\subsubsection{Copy-Synthesis (CS)}

As illustrated in the top of Fig.~\ref{fig:train}, copy-synthesis training aims to help the decoder generate high-fidelity audio segments while distilling the pitch information from the latent representation. Specifically, the generator seeks to reconstruct the input audio that cannot be distinguished by the discriminators while avoiding pitch information leakage to the latent representation, the pitch predictor aims to extract the pitch information from the latent representation, and the discriminator seeks to distinguish the ground truth audio and the predicted audio. Thus, we have: 
\begin{equation}
\begin{split}
    &L_\text{gen}^\text{CS} = L_\text{leak}(\hat{y}) + 15L_\text{mel}(\text{mel}, \hat{\text{mel}}) \\ 
    &\quad\quad + \sum_{m=1}^{M}[L_\text{adv}(G; D_{m}) + 2L_\text{feat}(G; D_m)], \\
    &L_\text{pred}^\text{CS} = L_\text{pitch}(y, \hat{y}), \\
    &L_\text{disc}^\text{CS} = \sum_{m=1}^{M}L_\text{adv}(D_{m}; G), \\
\end{split}
\end{equation}
where $L_\text{mel}$ is the multi-scale mel-spectrogram loss adapted from DAC~\cite{dac}, $D_m$ is the $m_\text{th}$ discriminator. $G$ is the Neurodyne. The $L_\text{adv}$ and $L_\text{feat}$ are the adversarial losses and the feature matching loss following HiFiGAN~\cite{HiFiGAN}.



\subsubsection{Inversion Cycle-Consistency (ICC)}

As illustrated in the middle of Fig.~\ref{fig:train}, inversion cycle-consistency training aims to enhance the pitch-manipulation ability of the model by implicitly creating paired in- and out-of-tuned data in the training process. The main idea is that the pitch-manipulated audio should still be able to be converted back, given the ground truth pitch. Under this scenario, the discriminator loss remains the same as the copy-synthesis training, while the generator and predictor losses become:
\begin{equation}
\begin{split}
    &L_\text{gen}^\text{ICC} = L_\text{leak}(\hat{y}) + L_\text{leak}(\hat{y}_\text{manipulated}) + 15L_\text{mel}(\text{mel}, \hat{\text{mel}}) \\ 
    &\quad\quad+ \sum_{m=1}^{M}[L_\text{adv}(G; D_{m}) + 2L_\text{feat}(G; D_m)], \\
    & L_\text{pred}^\text{ICC} = L_\text{pitch}(y, \hat{y}) + L_\text{pitch}(y_\text{manipulated}, \hat{y}_\text{manipulated}). \\
\end{split}
\end{equation}


\subsubsection{Composition Cycle-Consistency (CCC)}

As illustrated in the bottom of Fig.~\ref{fig:train}, composition cycle-consistency training ensures the audio manipulated to a specific pitch contour in one or two steps are identical. Under this scenario, there is no discriminator loss since there is no ground truth audio, while the generator and predictor losses become:
\begin{equation}
\begin{split}
    &L_\text{gen}^\text{CCC} = L_\text{leak}(\hat{y}_A) + L_\text{leak}(\hat{y}_{B_1}) \\
    &\quad\quad+ L_\text{leak}(\hat{y}_{B_2}) + 15L_\text{mel}(\hat{\text{mel}}_{B_1}, \hat{\text{mel}}_{B_2}), \\
    &L_\text{pred}^\text{CCC} = L_\text{pitch}(y_{A}, \hat{y}_{A}) + L_\text{pitch}(y_{B_1}, \hat{y}_{B_1}) \\
    &\quad\quad+ L_\text{pitch}(y_{B_2}, \hat{y}_{B_2}), \\
\end{split}
\end{equation}
where $B_1$ denotes modifying to the target pitch contour in one step, $A$ denotes the intermediate pitch contour in the two-step pitch manipulation, and $B_2$ denotes modifying to the target pitch contour in two steps via the intermediate pitch contour.


\subsection{Pitch Manipulation Strategy}

\begin{figure}[t]
    \centering
\includegraphics[width=\linewidth]{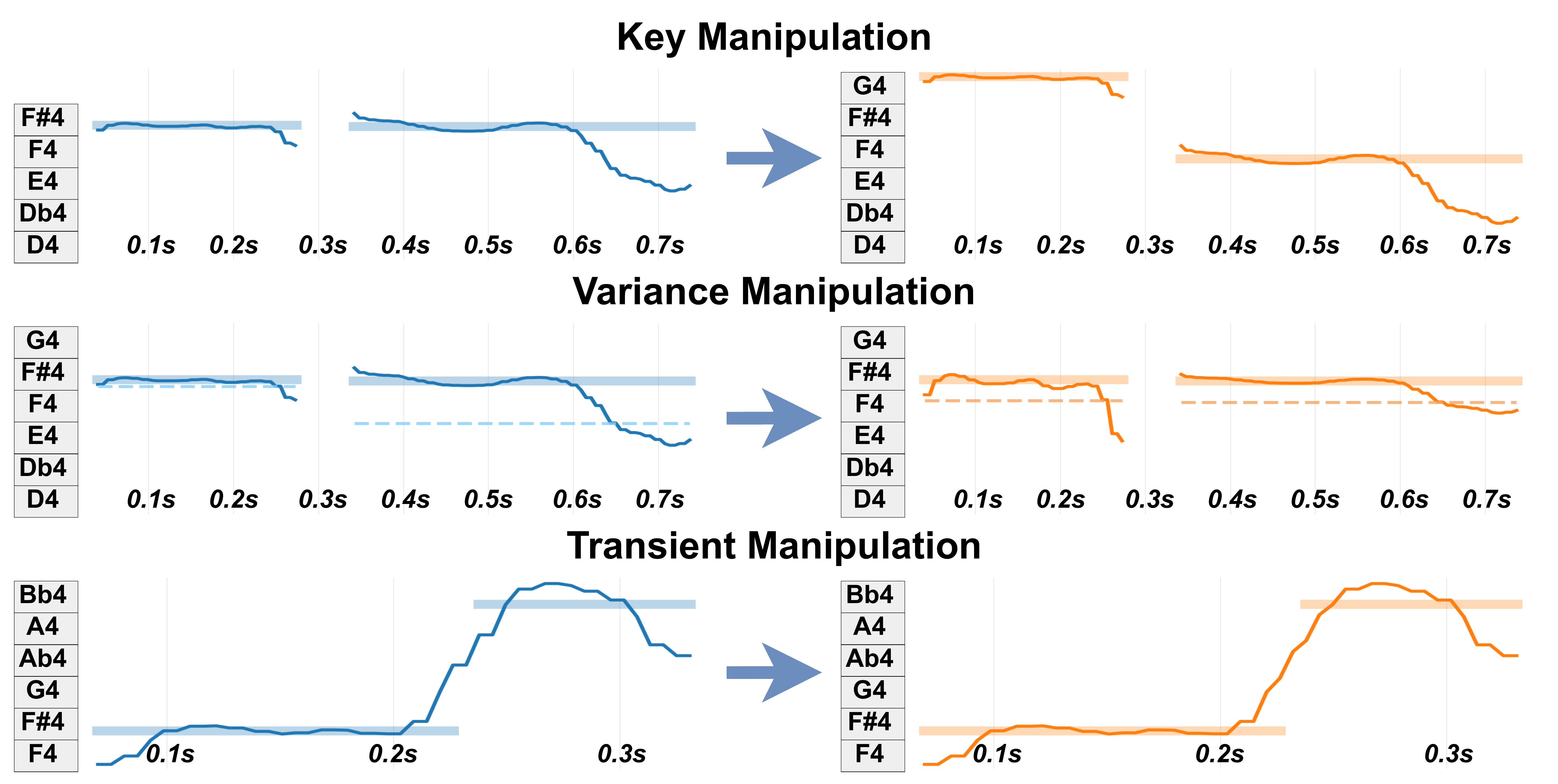}
    \vspace{-14pt}
    \caption{The pitch manipulation strategy.}
    \label{fig:shift}
    \vspace{-18pt}
\end{figure}

As illustrated in Fig.~\ref{fig:shift}, the pitch-manipulation strategy consists of three operations between the pitch contour and their center MIDI notes. The note value and rough segmentation are obtained from SOME\footnote{\url{https://github.com/openvpi/SOME}}, and the detailed segmentation is derived following CREPENotes~\cite{crepenote}. Specifically, 1) the key manipulation will randomly shift the pitch contour regarding each note up or down, 2) the variance manipulation will randomly shift the variance computed by subtracting the pitch contour and the note value up or down, 
3) the transient manipulation will randomly adjust the variance of the pitch contour between steady-state notes and the linear ramp between the notes.

\section{Experiments}

\begin{table*}[t]
\begin{center}
\caption{Global-key pitch manipulation results of different systems. The best and the second best results of every column (except those from Ground Truth) are \textbf{bold} and \underline{underlined}. The MUSHRA scores are within 95\% Confidence Interval (CI).}
\vspace{-7pt}
\label{tab:results-global-objective}
\setlength\tabcolsep{1.5pt}
\resizebox{\textwidth}{!}
{
\begin{tabular}{lccccccccccccccccccccc}

\toprule
\multirow{2}{*}{\textbf{System}} & \multicolumn{7}{c}{\textbf{F0RMSE ($\downarrow$)}} & \multicolumn{7}{c}{\textbf{Q-MUSHRA ($\uparrow$)}} & \multicolumn{7}{c}{\textbf{S-MUSHRA ($\uparrow$)}}\\ 
\cmidrule(lr){2-8} \cmidrule(lr){9-15} \cmidrule(lr){16-22}
& \textbf{-12} & \textbf{-6} & \textbf{-3} & \textbf{0} & \textbf{+3} & \textbf{+6} & \textbf{+12} & \textbf{-12} & \textbf{-6} & \textbf{-3} & \textbf{0} & \textbf{+3} & \textbf{+6} & \textbf{+12} & \textbf{-12} & \textbf{-6} & \textbf{-3} & \textbf{0} & \textbf{+3} & \textbf{+6} & \textbf{+12} \\
\midrule
Ground Truth & / & / & / & 0.0 & / & / & / & / & / & / & 93.0\scalebox{0.85}{$\pm$}2 & / & / & / & / & / & / & 93.0\scalebox{0.85}{$\pm$}2 & / & / & /\\ 
\midrule
 BigVGAN & / & / & / & \textbf{16.0} & / & / & / & / & / & / & 87.5\scalebox{0.85}{$\pm$}3 & / & / & / & / & / & / & 89.7\scalebox{0.85}{$\pm$}3 & / & / & / \\ 
\midrule
 WORLD & 20.2 & 24.8 & 28.1 & 31.7 & 37.8 & 46.4 & 96.3 & 42.3\scalebox{0.85}{$\pm$}5 & 52.7\scalebox{0.85}{$\pm$}5 & 59.1\scalebox{0.85}{$\pm$}6 & 75.7\scalebox{0.85}{$\pm$}5 & 65.7\scalebox{0.85}{$\pm$}5 & 63.8\scalebox{0.85}{$\pm$}5 & 55.0\scalebox{0.85}{$\pm$}6 & 41.5\scalebox{0.85}{$\pm$}5 & 57.2\scalebox{0.85}{$\pm$}5 & 67.3\scalebox{0.85}{$\pm$}5 & 83.2\scalebox{0.85}{$\pm$}5 & 72.9\scalebox{0.85}{$\pm$}5 & 68.4\scalebox{0.85}{$\pm$}5 & 59.8\scalebox{0.85}{$\pm$}5 \\
 TD-PSOLA & 29.0 & \underline{21.2} & \textbf{20.8} & \underline{20.0} & \textbf{28.6} & \underline{38.4} & \underline{90.2} & 42.3\scalebox{0.85}{$\pm$}5 & 55.4\scalebox{0.85}{$\pm$}6 & 69.0\scalebox{0.85}{$\pm$}4 & \textbf{90.0}\scalebox{0.85}{$\pm$}\textbf{2} & 71.1\scalebox{0.85}{$\pm$}4 & 60.8\scalebox{0.85}{$\pm$}5 & 58.3\scalebox{0.85}{$\pm$}5 & 43.6\scalebox{0.85}{$\pm$}5 & 59.1\scalebox{0.85}{$\pm$}5 & 73.4\scalebox{0.85}{$\pm$}4 & \textbf{91.8}\scalebox{0.85}{$\pm$}\textbf{2} & 78.8\scalebox{0.85}{$\pm$}4 & 65.3\scalebox{0.85}{$\pm$}5 & 62.4\scalebox{0.85}{$\pm$}5 \\
\midrule
 DiffPitcher & 32.0 & 39.4 & 45.9 & 52.8 & 65.1 & 83.6 & 163.0 & 25.5\scalebox{0.85}{$\pm$}4 & 36.6\scalebox{0.85}{$\pm$}4 & 44.5\scalebox{0.85}{$\pm$}5 & 57.2\scalebox{0.85}{$\pm$}6 & 44.8\scalebox{0.85}{$\pm$}5 & 41.5\scalebox{0.85}{$\pm$}5 & 26.5\scalebox{0.85}{$\pm$}4 & 32.5\scalebox{0.85}{$\pm$}5 & 47.9\scalebox{0.85}{$\pm$}5 & 61.0\scalebox{0.85}{$\pm$}5 & 75.1\scalebox{0.85}{$\pm$}5 & 63.4\scalebox{0.85}{$\pm$}5 & 53.5\scalebox{0.85}{$\pm$}6 & 39.5\scalebox{0.85}{$\pm$}6 \\
 SiFi-GAN & \underline{19.8} & 24.2 & 26.4 & 29.2 & 36.4 & 47.2 & 102.2 & 58.0\scalebox{0.85}{$\pm$}6 & \underline{66.6\scalebox{0.85}{$\pm$}4} & 78.0\scalebox{0.85}{$\pm$}4 & 89.2\scalebox{0.85}{$\pm$}3 & \underline{78.9\scalebox{0.85}{$\pm$}3} & \underline{73.7\scalebox{0.85}{$\pm$}6} & \textbf{65.9}\scalebox{0.85}{$\pm$}\textbf{6} & 51.3\scalebox{0.85}{$\pm$}6 & 65.0\scalebox{0.85}{$\pm$}5 & 79.6\scalebox{0.85}{$\pm$}3 & 90.2\scalebox{0.85}{$\pm$}3 & 80.0\scalebox{0.85}{$\pm$}4 & \underline{73.9\scalebox{0.85}{$\pm$}5} & \textbf{66.2}\scalebox{0.85}{$\pm$}\textbf{5} \\
 PC-NSF & 23.1 & 26.4 & 27.0 & 25.8 & 36.2 & 59.6 & 231.5 & \underline{59.4\scalebox{0.85}{$\pm$}6} & 66.1\scalebox{0.85}{$\pm$}5 & \underline{80.1\scalebox{0.85}{$\pm$}3} & \underline{89.8\scalebox{0.85}{$\pm$}3} & 78.9\scalebox{0.85}{$\pm$}4 & 72.3\scalebox{0.85}{$\pm$}5 & 51.9\scalebox{0.85}{$\pm$}6 & \underline{52.7\scalebox{0.85}{$\pm$}6} & \underline{66.2\scalebox{0.85}{$\pm$}4} & \underline{80.1\scalebox{0.85}{$\pm$}4} & \underline{90.6\scalebox{0.85}{$\pm$}2} & \textbf{81.8}\scalebox{0.85}{$\pm$}\textbf{4} & 73.6\scalebox{0.85}{$\pm$}5 & 56.9\scalebox{0.85}{$\pm$}6 \\
\midrule
 Neurodyne & \textbf{17.6} & \textbf{20.3} & \underline{22.1} & 24.0 & \underline{29.6} & \textbf{36.1} & \textbf{81.3} & \textbf{72.5}\scalebox{0.85}{$\pm$}\textbf{5} & \textbf{77.7}\scalebox{0.85}{$\pm$}\textbf{5} & \textbf{81.1}\scalebox{0.85}{$\pm$}\textbf{3} & 88.4\scalebox{0.85}{$\pm$}4 & \textbf{79.0}\scalebox{0.85}{$\pm$}\textbf{4} & \textbf{76.3}\scalebox{0.85}{$\pm$}\textbf{4} & \underline{62.0\scalebox{0.85}{$\pm$}6} & \textbf{64.4}\scalebox{0.85}{$\pm$}\textbf{6} & \textbf{74.9}\scalebox{0.85}{$\pm$}\textbf{5} & \textbf{80.6}\scalebox{0.85}{$\pm$}\textbf{3} & 90.5\scalebox{0.85}{$\pm$}3 & \underline{80.1\scalebox{0.85}{$\pm$}4} & \textbf{75.4}\scalebox{0.85}{$\pm$}\textbf{5} & \underline{64.4\scalebox{0.85}{$\pm$}6} \\
\bottomrule

\end{tabular}
} 
\vspace{-24pt}
\end{center}
\end{table*}

We evaluate the effectiveness of our proposed system in two settings.
We apply global-key~\cite{sifigan} and template-based~\cite{diffpitcher} pitch manipulation to compare the robustness and evaluate the application in real-world scenarios.
In global-key pitch manipulation, utterances will be globally manipulated by a specific amount of semitones. In template-based pitch manipulation, an out-of-tune audio segment will be adjusted based on an in-tune reference. The audio samples are available on our demo page\footnote{\label{footnote:demopage}\url{https://www.yichenggu.com/Neurodyne/}}.

\subsection{Experiment Setup}

\quad \hskip0.6em\relax \textbf{Datasets}: Following~\cite{emilia, emilia-journal, singnet, sslbus}, we collected all publicly available academic datasets to obtain a diversified large-scale data mixture, which is illustrated in Table~\ref{tab:datasets}. We use the paired recordings in PopButFy~\cite{popbutfy} for evaluating template-based pitch manipulation, where 200 utterance pairs with significant pitch differences were selected. We randomly selected 1\% samples from the remaining dataset to form the test set for global-key pitch manipulation. All the remaining samples were used in training, resulting in a 580-hour data mixture.

\begin{table}[t]
    \centering
    \caption{Statistics of the singing voice datasets.}
    \label{tab:datasets}
    \vspace{-3pt}
    \resizebox{\linewidth}{!}{
        \begin{tabular}{cccc}
            \toprule
            \textbf{Dataset} & \textbf{Dur.~(hour)} & \textbf{Style} & \textbf{Lang.} \\
            \midrule
            NUS-48E~\cite{NUS48E} & 2.8 & Children/Pop & ZH \\
            Opera~\cite{opera} & 2.6 & Opera & IT/ZH \\
            VocalSet~\cite{VocalSet} & 8.8 & Opera & EN \\
            CSD~\cite{csd} & 4.6 & Children & EN/KO \\
            PJS~\cite{pjs} & 0.5 & Pop & JA \\
            NHSS~\cite{NHSS} & 4.1 & Pop & EN \\
            OpenSinger~\cite{OpenSinger} & 51.8 & Pop & ZH \\
            Kiritan~\cite{kiritan} & 1.2 & Pop & JA \\
            KiSing~\cite{kising} & 0.9 & Pop & ZH \\
            PopCS~\cite{diffsinger} & 5.9 & Pop & ZH \\
            M4Singer~\cite{M4Singer} & 29.7 & Pop & ZH \\
            PopBuTFy~\cite{popbutfy} & 30.7 & Pop & EN \\
            Opencpop~\cite{Opencpop} & 5.2 & Pop & ZH \\
            SingStyle111~\cite{SingStyle} & 12.8 & \makecell{Children/Folk/Jazz \\ Opera/Pop/Rock} & EN/IT/ZH \\
            GOAT~\cite{goat} & 4.5 & Opera & ZH \\
            ACESinger~\cite{ACESinger} & 321.8 & Pop & EN/ZH \\
            GTSinger~\cite{gtsinger} & 96.8 & \makecell{Folk/Jazz \\ Opera/Pop} & \makecell{ZH/EN/JA \\ KO/RU/ES \\ FR/DE/IT} \\
            \bottomrule
        \end{tabular}
    }
    \vspace{-16pt}
\end{table}

\textbf{Preprocessing}: We resampled all the training data to 44.1kHz. We use the TorchFCPE~\footnote{\label{footnote:fcpe}\url{https://github.com/CNChTu/FCPE}} to extract the F0. For computing the input features, we use an FFT size of 2048, hop size of 512, window length of 2048, Mel filters of 128, and Mel-cepstral coefficient orders of 39. The mel-spectrogram is normalized in log-scale with values $\leq$ 1e-5 clipped to 0.

\textbf{Training}: All the models are trained using the AdamW~\cite{adamw} optimizer with $\beta _ {1} = 0.9$, $\beta _ {2} = 0.998$, and an learning rate of 1e-4, and the exponential decay scheduler with a factor $\gamma = 0.999996$. All the experiments are conducted on 8 MI250X GPUs with the maximum available batch size for 1M steps.


\textbf{Configurations}: For baseline systems, BigVGAN~\cite{bigvgan} is used as the benchmark for copy-synthesis. We use WORLD~\cite{world} and TD-PSOLA~\cite{psola} as the DSP-based baselines. We use Diffpitcher~\cite{diffpitcher}, SiFi-GAN~\cite{sifigan}, and HarmonicNet~\cite{pcnsf} as the DL-based baselines. For SiFi-GAN, we use the 44.1\,kHz configuration\footnote{\url{https://github.com/tonnetonne814/SiFi-VITS2-44100-Ja}}. For HarmonicNet, we use the implementation open-sourced by the OpenVPI team\footnote{\label{footnote:openvpi}\url{https://github.com/openvpi/SingingVocoders}}, denoted as PC-NSF. For Neurodyne, the encoder adjusts the DAC~\cite{dac} with an output dimension of 128; the pitch predictor adjusts the TorchFCPE\footnoteref{footnote:fcpe} with the latent representation as the input; the decoder applies $2\times$ oversampling to the NSF-HiFiGAN~\cite{SVCC}\footnoteref{footnote:openvpi}; the multi-scale oscillator uses sampling rates of [5512.5, 11025, 22050, 44100]; the MS-SB-CQT Discriminator~\cite{cqt} computes 10 octaves to conform the 44.1\,kHz training, with other discriminators unchanged.

\subsection{Evaluation Metrics}

\quad \hskip0.6em\relax \textbf{Objective Evaluation:} 
We use the Amphion~\cite{amphion} toolkit for objective evaluation. We investigate objective metrics focusing on F0 accuracy, audio quality, and singer similarity. For F0 accuracy, we use the F0 Root Mean Square Error (F0RMSE) and F0 Pearson Correlation Coefficient (FPC)~\cite{SVCC, contentsvc}. We use a pre-trained MOS predictor~\cite{sheet} to obtain the predicted audio quality score (MOS-Pred); we use all open-sourced SSL models from~\cite{singssl} to compute the singer similarity (SIM-O) and report their average score for the ablation study.

\textbf{Subjective Evaluation:} We use a MUSHRA-like test to evaluate the audio quality and singer similarity subjectively. A total of 35 and 10 utterances will be assessed in each setting individually. Listeners were asked to give quality and similarity scores (denoted as Q- and S-MUSHRA) between 1 and 100 for global-key or template-based pitch-manipulated utterances from different systems given the ground truth or in- and out-of-tune audio as the reference. We invited 15 volunteers who are experienced in audio generation with the ability to distinguish out-of-tune segments to attend the evaluation. 


\begin{table}[t]
\begin{center}
\caption{Template-based pitch manipulation results of different systems. The best and the second best results of every column (except those from Ground Truth) are \textbf{bold} and \underline{underlined}. The MUSHRA scores are within 95\% Confidence Interval (CI).}
\vspace{-6pt}
\label{tab:results-template}
\setlength\tabcolsep{3pt}
\resizebox{\linewidth}{!}{
\begin{tabular}{lcccc}

\toprule
\textbf{System} & \textbf{FPC ($\uparrow$)} & \textbf{F0RMSE ($\downarrow$)} & \textbf{Q-MUSHRA ($\uparrow$)} & \textbf{S-MUSHRA ($\uparrow$)} \\
 \midrule
 WORLD & 0.951 & 40.7 & 61.9\scalebox{0.85}{$\pm$}3 & 67.8\scalebox{0.85}{$\pm$}3 \\
 TD-PSOLA & \underline{0.970} & \underline{30.6} & 67.3\scalebox{0.85}{$\pm$}3 & 71.1\scalebox{0.85}{$\pm$}3 \\
 \midrule
 DiffPitcher & 0.868 & 77.2 & 41.9\scalebox{0.85}{$\pm$}3 & 53.9\scalebox{0.85}{$\pm$}4 \\
 SiFi-GAN & 0.969 & 31.3 & 77.5\scalebox{0.85}{$\pm$}2 & 78.4\scalebox{0.85}{$\pm$}2 \\
 PC-NSF & 0.966 & 34.8 & \underline{78.2\scalebox{0.85}{$\pm$}2} & \underline{78.7\scalebox{0.85}{$\pm$}2} \\
 \midrule
 Neurodyne & \textbf{0.978} & \textbf{27.1} & \textbf{78.3}\scalebox{0.85}{$\pm$}\textbf{2} & \textbf{78.8}\scalebox{0.85}{$\pm$}\textbf{2} \\
\bottomrule
\end{tabular}
}
\vspace{-26pt}
\end{center}
\end{table}

\subsection{Global-Key Pitch Manipulation}
\label{sec:global}

The global-key pitch manipulation evaluation results are illustrated in Table~\ref{tab:results-global-objective}. It can be observed that 1) neural-vocoder-based systems perform significantly better than the two-stage synthesis and DSP-based systems, confirming the effectiveness of adversarial training with audio-level losses;
2) in copy-synthesis, Neurodyne performs better subjectively compared with BigVGAN, which shows the effectiveness of the less information loss brought by the adversarial representation learning, making it easier to generate high-fidelity audio;
3) in different pitch-manipulation scenarios, Neurodyne performs better than the baseline systems with smaller F0RMSE and higher quality and similarity MUSHRA scores. This illustrates the effectiveness of feature disentanglement 
and robustness, which is brought by cycle-consistency training and adversarial representation learning. Specifically, the network will automatically adjust the latent representation regarding the modified F0 to avoid timbre inconsistency brought by the inaccurate disentanglement. Meanwhile, extreme pitch-manipulation scenarios are seen in the training stage due to the implicit data construction, which leads to improved F0 accuracy and audio quality. 


\subsection{Template-Based Pitch Manipulation}
\label{sec:template}

The template-based pitch manipulation evaluation results are illustrated in Table~\ref{tab:results-template}. Our system performs better in subjective synthesis quality, singer similarity, and objective pitch accuracy in both absolute value and relative trajectory, which shows the effectiveness of our model architecture, training schemes, and pitch-manipulation strategies.

To further illustrate the effectiveness of our proposed system, we also compared Neurodyne 
with two SOTA commercial softwares. Specifically, we manually selected 4 Chinese and 4 English representative paired samples and manually conducted pitch manipulation. 
We post the audio samples on our demo page~\footnoteref{footnote:demopage}, and it can be heard that our model performs equivalently or even better compared with these commercial plugins.

\subsection{Ablation Study}
\label{sec:ablation}

\begin{figure}[t]
    \centering
    \vspace{-24pt}\includegraphics[width=\linewidth]{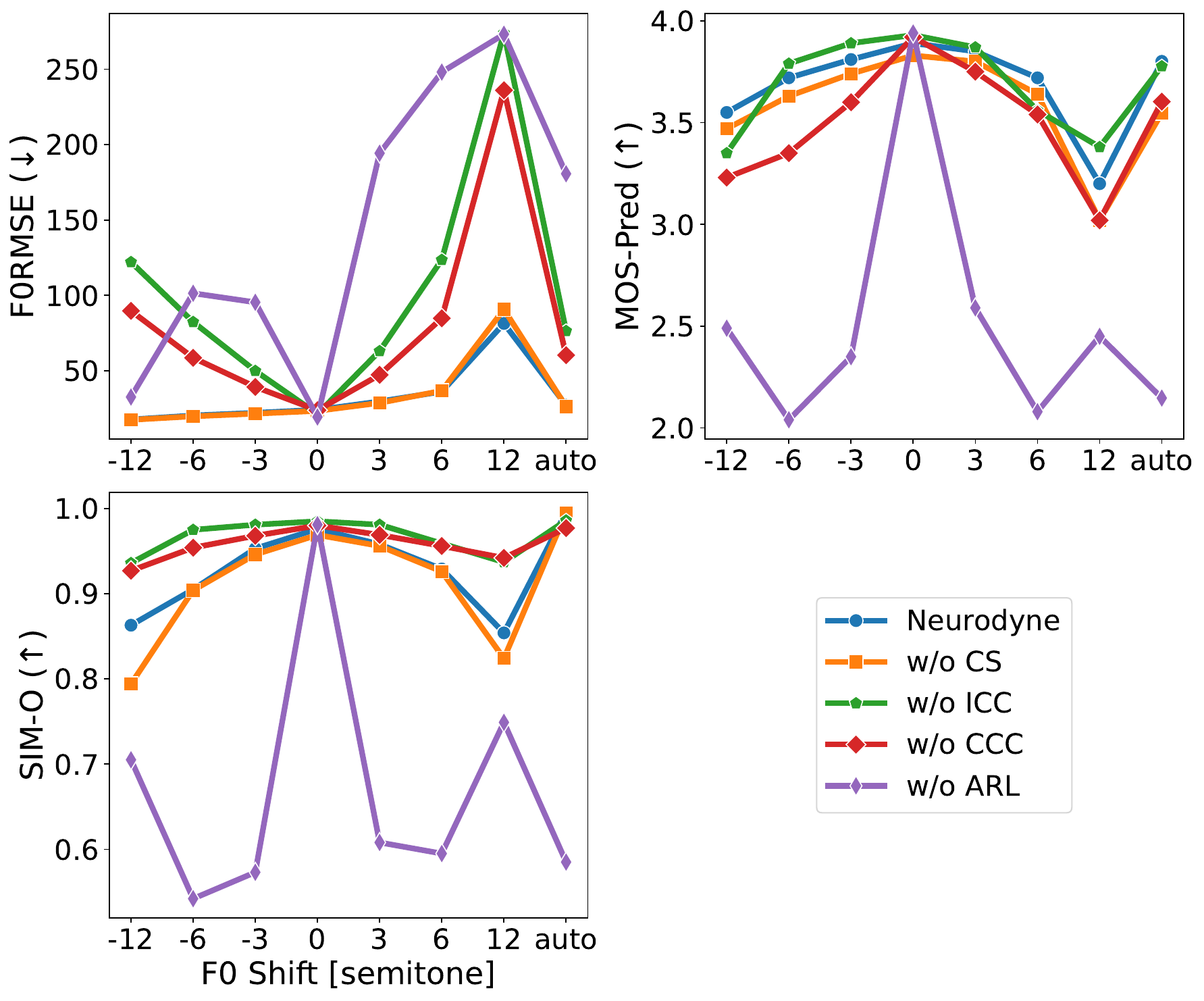}
    \caption{The evaluation results of the ablation study. ``CS'' means copy-synthesis, ``ICC'' means inversion cycle-consistency, ``CCC'' means composition cycle-consistency, and ``ARL'' means adversarial representation learning.}
    \label{fig:ablation}
    \vspace{-12pt}
\end{figure}


We conducted an ablation study to illustrate the effectiveness of cycle-consistency training and adversarial representation learning, as shown in Fig.~\ref{fig:ablation}. Firstly, the system without copy-synthesis training has worse synthesis quality and singer similarity scores, showing it is necessary to conduct copy-synthesis training to maintain the model performance. Moreover, the system without inversion or composition cycle-consistency training will fail to manipulate pitch according to the large F0RMSE values. This is because the constrain for the overall system is loosened, and the decoder will ignore the given pitch condition and utilize the pitch information leaked from the encoder to the latent representation instead, generating either audio with the original pitch contour or unnatural audio comprising singing voices in both pitches (original and shifted). Lastly, the system without adversarial representation learning will only generate reasonable audio in the copy-synthesis scenario since the pitch information is all encoded in the latent representation. The model will be confused when the pitch condition is different, thus generating metallic noises, as illustrated in the unreasonably large F0RMSE, MOS-Pred, and SIM-O values.

\section{Conclusion}

This paper introduces Neurodyne, a novel pitch manipulation system that utilizes adversarial representation learning and cycle-consistency training. It obtains optimized pitch-independent features that avoid the artifacts brought by source-filter-model-based 
disentanglement. It implicitly creates paired in- and out-of-tune training data to enhance the robustness in different pitch manipulation scenarios. The evaluation results in both global-key and template-based pitch manipulation demonstrate that our proposed system outperforms existing ones regarding audio quality, pitch accuracy, and singer similarity.

\newpage

\section{Acknowledgement}

We acknowledge the computational resources provided by the Aalto Science-IT project. We acknowledge the EuroHPC Joint Undertaking for awarding this project access to the EuroHPC supercomputer LUMI, hosted by CSC (Finland) and the LUMI consortium through a EuroHPC Regular Access call. This work is also supported by the 2023 Shenzhen stability Science Program, the Program for Guangdong Introducing Innovative and Entrepreneurial Teams (Grant No. 2023ZT10X044), and the Shenzhen Science and Technology Program (ZDSYS20230626091302006)

{
\bibliographystyle{IEEEtran}
\bibliography{mybib}
}

\end{document}